\documentclass[11pt]{article}
\pdfoutput=1
\usepackage{epsfig}
\usepackage{graphicx}
\usepackage{multicol}
\usepackage{amsmath,amssymb}
\numberwithin{equation}{section}
\usepackage{cite}

\newcommand{\bea}{\begin{eqnarray}}
\newcommand{\eea}{\end{eqnarray}}
\newcommand{\be}{\begin{equation}}
\def\bel#1{\begin{equation} \label{#1}}
\newcommand{\ee}{\end{equation}}
\newcommand{\ba}{\begin{align}}
\newcommand{\ea}{\end{align}}
\newcommand{\comments}[1]{}

\def\pref#1{(\ref{#1})}

\newcommand\czt{{\mathbb{C}^{3}/ \mathbb{Z}_{3}}}
\def\op{\oplus}
\begin{document}
\begin{titlepage}
\today \hfill{DAMTP-2011-101}
\vskip 4 cm
\begin{center}
{\Large \bf  
Symmetry Breaking Bulk Effects in Local D-brane Models
}
\vskip 1.5cm  
{ 
Anshuman Maharana
\let\thefootnote\relax\footnotetext{electronic address: {$\mathtt{am794@cam.ac.uk} $}}
}
\vskip 0.9 cm
{\textsl{
DAMTP, University of Cambridge, \\
Wilberforce Road, Cambridge,\\
CB3 0WA, UK.\\}
}
\end{center}

\vskip 0.6cm

\begin{abstract}
We  study symmetrybreaking effects in local D-brane models that arise as a result of
 compactification, focusing on models constructed on $\mathbb{C}^{3}/ \mathbb{Z}_{3}$. 
 Zero-modes  of the Lichnerowicz operator  in cone-like geometries have a power law behaviour; thus the leading symmetry breaking effects are captured by the modes with the lowest scaling dimension which transform non-trivially under the isometry group. Combining this with the fact that global symmetries in local models are gauged
upon compactification we determine the strength and form of the leading operators
responsible for the symmetry breaking.
\noindent

\end{abstract}

\vspace{3.0cm}

\end{titlepage}
\pagestyle{plain}
\setcounter{page}{1}
\newcounter{bean}
\baselineskip18pt
%


\section{Introduction}

   The bottom-up approach \cite{bottomup,otherbot, Verlinde, realistic, toric, dpt, Vjay, fone, FthSym, warpedchiral} is a promising  direction for string phenomenology. In this approach the
  Standard Model degrees of freedom arise from a D-brane construction
  which is localised in the extra dimensions. Various
  properties of the Standard Model sector such as matter content, gauge couplings
  and Yukawa couplings can be computed in the local setting, without knowledge
  of the global aspects of the compactification.  A modular approach is possible.
  After having constructed a realistic Standard Model sector in a local geometry one can then attempt to
  embed the model in a compactifiction. At the later stage one has to consider global
  aspects such as tadpole cancellation, moduli stabilisation,  supersymmetry breaking
  and cosmology.
  
        Isometries in the geometry  imply global symmetries
    in the local model.  Such symmetries can have  phenomenologically interesting implications. For instance, it was shown in \cite{radfermion}
 that   for the models constructed in \cite{realistic,toric}
 isometries of the local geometry imply that fermions in the lightest generation are
massless at tree-level. 

   The fate of such symmetries once the local model is embedded in a compactification
was discussed in \cite{gaugelocal}.  The effect of compactification is to couple the model to gravity; theories of quantum gravity cannot  possess exact continuous global symmetries \cite{BanksDix, BanksSei, Hellsha}. It was argued in
\cite{gaugelocal} that once  compactification effects are taken into account  global symmetries in local models are gauged and the associated gauge field
is a closed string (bulk) mode.  Furthermore, compact Calabi Yaus  with $SU(3)$ holonomy
do not have any isometries. This implies that the gauge symmetry is spontaneously broken by
the bulk. These effects are communicated to the open string sector via
their interactions with the bulk modes. Once the bulk modes are integrated out one
is left with an open string sector which has an approximate global symmetry, with the
closed string sector providing the ``flavon'' vevs which parametrise the symmetry breaking.
The strength of symmetry breaking necessarily scales as an inverse power of the compactification volume; as  in the infinite volume limit Newton's constant vanishes and the global symmetry is restored.  Couplings in the Standard Model sector which vanish in the non-compact analysis due to the presence of the
global symmetry can take on values which scale as inverse powers of the volume; if the
volume of the compactification is large  one can generate parametrically small couplings in the Standard Model sector.

 The $\czt$ geometry provides an attractive starting point  \cite{bottomup} for local model building. Models
on complex cones over the lower del Pezzo surfaces \cite{realistic, toric, dpt} are closely related, in order to
obtain the Standard Model gauge groups at low energies one has to Higgs these models
to $\czt$.  The geometry has a 
$SU(3) \times U(1)$ isometry.  In this paper we shall focus on $\mathbb{C}^{3}/ \mathbb{Z}_{3}$ (either singular or resolved); and compute the volume dependence and form of the leading operators that break the $SU(3) \times U(1)$  symmetry as a result of compactification.  We shall do so by determining the leading deviations in the metric which
break the  isometries once the  geometry is glued on to a bulk to form
a compact space. These can be obtained  by carrying out a small fluctuation analysis in the local geometry, along the
 lines of \cite{AAB, holo2, Gandhi}. For simplicity, we will model the effect of compactification on the local geometry via a deformation which preserves the Ricci flatness condition. However, general supersymmetry-breaking deformations will not preserve Ricci flatness, and a systematic study of supersymmetry-breaking effects requires the analysis of other modes such as the fluxes and their backreaction on the metric. We hope to return to these issues in the future.   

                In the region where the deformation is small
compared to the background metric, a Ricci flat  deformation can be written as a series
in zero-modes of the Lichnerowicz operator constructed from the local geometry.
For cone-like geometries zero-modes of the Lichnerowicz operator have angular
dependence given by symmetric  tensor harmonics on the base space and a power
law radial behaviour. Thus,
\bel{expan}
   \delta g_{ij} = \sum_{I} c_{I} Y^{I}_{ij} (\psi) r^{ \Delta_{I}}
\ee
where $Y^{I}_{ij}$ are harmonics on the base, $\Delta_{I}$ 
the radial scaling associated with the harmonic $Y^{I}_{ij}$     and $c_{I}$ Fourier
coefficients. Modes with positive $\Delta_{I}$ grow in the asymptotic end of
the cone and fall off in the tip region: these are  induced
as a result of compactification \footnote{Modes with negative $\Delta_{I}$ have profiles
that vanish in the asymptotic region and are to be associated with blow-up modes.}. In the bulk of the compactification the 
isometries are badly broken, the deformed non-compact metric smoothly connects onto this
``generic'' metric; we therefore will assume  that  in the asymptotic region of the cone
all  terms in the sum \pref{expan} are of equal magnitude. Given this, the power law
behaviour in the radial direction implies that  the perturbation  in the vicinity of the tip of the cone
is well approximated by the modes with the
smallest positive values of $\Delta_{I}$. Thus in order to obtain the leading effects that
break a certain isometry one has to isolate the harmonic $Y^{I}_{ij}$ in the
expansion \pref{expan} with the smallest positive value of $\Delta_{I}$ which is not a singlet
under the action of the isometry group. 

Once we have isolated the relevant harmonics   we shall construct the lowest dimension operators  that involve
both the Standard Model sector and the symmetry breaking harmonics (flavons). These operators have to be gauge invariant under the action of the isometry group; this
will allow us to enumerate the form of the symmetry breaking operators that arise in the open string
sector. 
\section{Perturbations and Breaking of Isometries on $\czt$ }
\label{secpiso}
 The space   $\mathbb{C}^{3} / \mathbb{Z}_{3}$ is obtained from $\mathbb{C}^{3}$ by making the identification $z^{i} \equiv e^{2 \pi i /3} z^{i}$.  The  space inherits the K\"ahler potential of $\mathbb{C}^{3}$,
$ K = z^{i} \bar{z}_{\bar{i}} = r^{2} $. The identification breaks the $SO(6)$ isometry of $\mathbb{C}^{3}$
to $U(3) \cong SU(3) \times U(1)$. Under the action of the isometry group  the complex coordinates $z^{i}$
 transform in the fundamental, $z^i \to U^{i}_{\phantom{i}j} z^{j}$.

     The origin is a fixed point of the orbifold action and is singular. The singularity can be resolved \cite{Lutken} by modifying
the K\"ahler potential to
\bel{lut}
 r^2 K'(r^{2}) = (r^6+r_0^6)^{1/3}.
\ee
The metric on the space then becomes
\be
    g_{m \bar{n}} = { ( r^{6} + r_{0}^{6} )^{1/3} \over r^{2} } \delta_{m \bar{n}}
 -   { r_{0}^{6} \over  {r^{4} ( r^{6} + r_{0}^{6} )^{2/3}}  } z_{m} z_{\bar{n}},
\ee
the locus $r=0$ now corresponds a finite size $\mathbb{P}^2$ with the canonical Fubini Study metric of radius
$r_0$.

\subsection*{Perturbation Analysis}

 In order to discuss perturbations it is useful to consider $\czt$ as a cone with base $S^{5}/ \mathbb{Z}_{3}$. Static perturbations of general CY cones
\be
\label{backg}
   ds_{6}^{2} = g_{mn} dy^{m} dy^{n} = dr^{2} + r^{2} ds^{2}_{{\cal{B}}_{5}} = dr^{2} + r^{2} \tilde{g}_{ij} d \Psi^{i} d \Psi^{j}
\ee
were discussed in  \cite{Gandhi}, the results of which we now review. For metric perturbations, $\delta g_{mn} = h_{mn}$,
linearized Einstein equations are given by $\Delta_{K} h_{mn} = 0 $, where $\Delta_{K}$ is the Lichnerowicz operator constructed from the metric $g_{mn}$. Working in transverse  gauge with respect to the base,
$\tilde{\nabla}^{i} {h}_{ij} = \tilde{\nabla}^{i} {h}_{ir} = 0$; Einstein equations imply ${h}_{ri} = h_{rr} = 0.$ The
angular components of the metric are required to take the form
\be
  h_{ij}  =  \sum_{I_t}\,\phi^{I_t}(r)Y^{I_t}_{ij}(\Psi),
\ee
with the index $I^t$  running over the symmetric traceless two tensor harmonics on the base ${{\cal{B}}_{5}}$.
These are eigenfunctions of the Lichnerowicz operator constructed from the base metric $\tilde{g}_{ij}$.
 The functions $\phi^{I_{t}}(r)$ are solutions of the differential equation
\be
\bigg( \partial_{r}^{2} - { {4} \over {r} } \partial_{r}  + { {4 - \lambda^{I_{t}}} \over r^{2} } \bigg) \phi^{I_{t}} = 0,
\ee
where $\lambda^{I_t}$ is the eigenvalue of the harmonic $Y^{I_t}_{ij}$.  The general solution of which is
\be
  \phi^{I_{t}}(r) =  a^{I_{t}}_{+} r^{+ \Delta}  + a^{I_{t}}_{-} r^{- \Delta} 
\ee
with
\bel{scale}
 \Delta  =   \sqrt{  \lambda^{I_{t}} - 4}
\ee
where $ a^{I_{t}}_{\pm}$ are integration constants, fixed by the boundary conditions.
   
\subsection*{Tensor Harmonics on $S^{5} / \mathbb{Z}_3$}
\label{secharmo}
 The tensor harmonics on $S^{5} / \mathbb{Z}_3$ can be obtained from those on $S^{5}$ by carrying out
an orbifold projection.  We  begin by discussing some basic properties of tensor
harmonics on d-spheres (see for e.g. \cite{KodIshi}). Tensor harmonics on the d-sphere can be constructed by considering functions of the form
\bea
  Y^{l}_{ij} = C^{l}_{A_1.....A_{l}; B_1B_2} \Omega^{A_1}.... \Omega^{A_2} \hat{D}_{i} \Omega^{B_1} \hat{D}_{j} \Omega^{B_2}, {\text{\phantom{aa}}} &&l \geq 2 \cr
  && A_1,.....,A_{l},B_1,B_2 = 1 ... d+1
 \eea
where $\Omega^{A}$ are homogeneous coordinates on the d-sphere $\Omega^{A}. \Omega^{A} =1$ and
$\hat{D}$ denotes the covariant derivative compatible with the d-sphere. $C^{l}_{A_1.....A_{l}; B_1B_2}$
is a constant tensor which is symmetric and traceless with respect to the indices $A_{i}$ and $B_j$ and satisfies
$C_{(A_1.....A_{l}; \phantom{B} A_{l+1})}^{l\phantom{A_1.....A_{l};}B} =C_{(A_1.....A_{l};  A_{l+1})}^{l\phantom{A_1.....A_{l}A_{l+1};}B} = 0$. For a fixed value of $l$ the harmonics transform as an irreducible representation
of $SO(d+1)$. For the five sphere the  associated highest weight state  has Dynkin label
$(l-2,2,2)$. The eigenvalue for the Lichnerowicz operator is
\be
  \lambda^{I_{t}}  =  {  l(l +4) + 8  } \ .
\ee
Equation  \pref{scale} then gives
\be
   \Delta = l+2 \ .
\ee

 In  order to obtain the harmonics on $S^{5}/ \mathbb{Z}_{3}$ we decompose the 
$SO(6)$ representations associated with the harmonics of $S^{5}$ under the maximal subgroup $SU(3) \times U(1)$ and project to states that have vanishing U(1) charge \footnote{We normalise the $U(1)$ charge so that the
vector of $SO(6)$  decomposes into the fundamental and anti-fundamental of $SU(3)$ with charge $+1$ and $-1$. } modulo 3. Table \ref{tab} lists
the $SU(3)$ representation and U(1) charge of  the harmonics 
for $l \leq 4$.  Details of the decomposition are given in the appendix.
\begin{center}
\begin{table}
\caption{Tensor Harmonics on $S^{5} / \mathbb{Z}_3$ for  ${l} \leq 4$.}
\vspace{0.3cm}
\begin{center}
  \begin{tabular}{ | c | p{4.3cm} | }
    \hline
    $l$ &   $SU(3)$ Representations and $U(1)$ charge \\ \hline
    2 &   $(0,0)^{0},(1,1)^{0},(2,2)^{0}$ \\   \hline
    3&     $(1,1)^{-3}, (3,0)^{-3}, (2,2)^{-3}$       
          $(1,1)^{+3}, (0,3)^{+3}, (2,2)^{+3}$ \\ \hline
    4 &      $(1,1)^{0}, (3,0)^{0}, (0,3)^{0},$  $(2,2)^{0}, (1,4)^{0}, (4,1)^{0},$  $(3,3)^{0}, (1,2)^{-6}, (2,1)^{+6}$ \\ \hline 
\end{tabular}
\end{center} 
\label{tab}
\end{table}
\end{center}

\subsection*{Strength of Isometry Breaking Effects}

     As discussed in the introduction, in the vicinity of the tip of the cone the leading  isometry
breaking effects are captured by the harmonics which have the lowest (positive) scaling dimension 
and transform non-trivially under the action of the isometry group \footnote{The singlet at $l=2$  with $\Delta=-4$ is related to the
resolution \pref{lut}. The associated mode with a vanishing profile in the asymptotic end has a fall off $r^{-4}$, i.e.
is suppressed by $r^{-6}$ relative to the background metric. This is in keeping with the small $r_{0}$ expansion of
\pref{lut}. }. In associating the effects of compactification only with modes  with positive 
scaling, we are assuming that there are no compactification effects that grow 
in the small r region. In general such effects can be present - the 
analogue of relevant deformations in AdS/CFT examples. The presence of such 
modes indicates that the small r geometry is unstable to certain 
deformations of the bulk. One would then like to find symmetry reasons to 
prevent these unstable modes to be turned; so that the geometry in small r 
(location of the brane set up) is stable. For the $SU(3)$ generators, 
Table \ref{tab} shows that these harmonics are at $l=2$ ($\Delta = 4$) with $SU(3)$ Dynkin label (1,1) and
(2,2). The background metric \pref{backg} has angular components which  scale as $r^{2}$; thus in the
 vicinity of the tip of the cone $\Delta =4$ deformations are suppressed by a factor  $ (r/ r_{\textrm{asym}})^2$ relative to the background  (where $r_{\textrm{asym}}$ is the value of the radial coordinate
at which the cone is glued  on to the bulk). For an open string sector which is localised on a $\mathbb{P}^{2}$ resolving the singularity at $r=0$, the 
 relevant scale at small $r$ is the resolution radius $r_{0}$. In the absence of any anisotropies in the compactification $r_{\textrm{asym}}$ is of the same scale as the compactification radius. Thus compactification effects that break the $SU(3)$ symmetry
 are  of magnitude 
 \be
   \epsilon \sim \bigg( { r_{0} \over r_{\textrm{asym}} } \bigg)^{2} \sim  \bigg( { \tau_{s} \over \tau_{b} } \bigg)^{1/2}.
\ee
Where $\tau_{s}$ and $\tau_{b}$ are the K\"ahler moduli associated with the blow-up mode and the volume modulus.
For models on the singular locus the relevant scale at small $r$ is the string scale, thus
\be
\label{flv}
   \epsilon \sim {1 \over {{\cal{V}}^{1/3}}} \ .
\ee
Where ${\cal{V}}$ is the volume of the compactification in string units.

  The $U(1)$ symmetry is unbroken at $l=2$; it is broken by $l=3$ modes with charge $\pm3$.
The associated suppression factor is
\be
  \delta \sim { 1 \over { {\cal{V}}^{1/2}}} \ .
\ee
 Thus at large volume  the $SU(3)$ and $U(1)$ symmetry breaking parameters
 are hierarchically separated.

\section{Symmetry Breaking Operators for D3-branes at $\czt$}
 
   In this section we shall consider D3-branes at a $\czt$ singularity and determine the
form of the symmetry breaking operators in the open string sector that are induced as a result
of compactification. Our guiding principle will be gauge invariance under the flavon (isometry) group.

\begin{figure}
\begin{center}
\includegraphics[height=55mm]{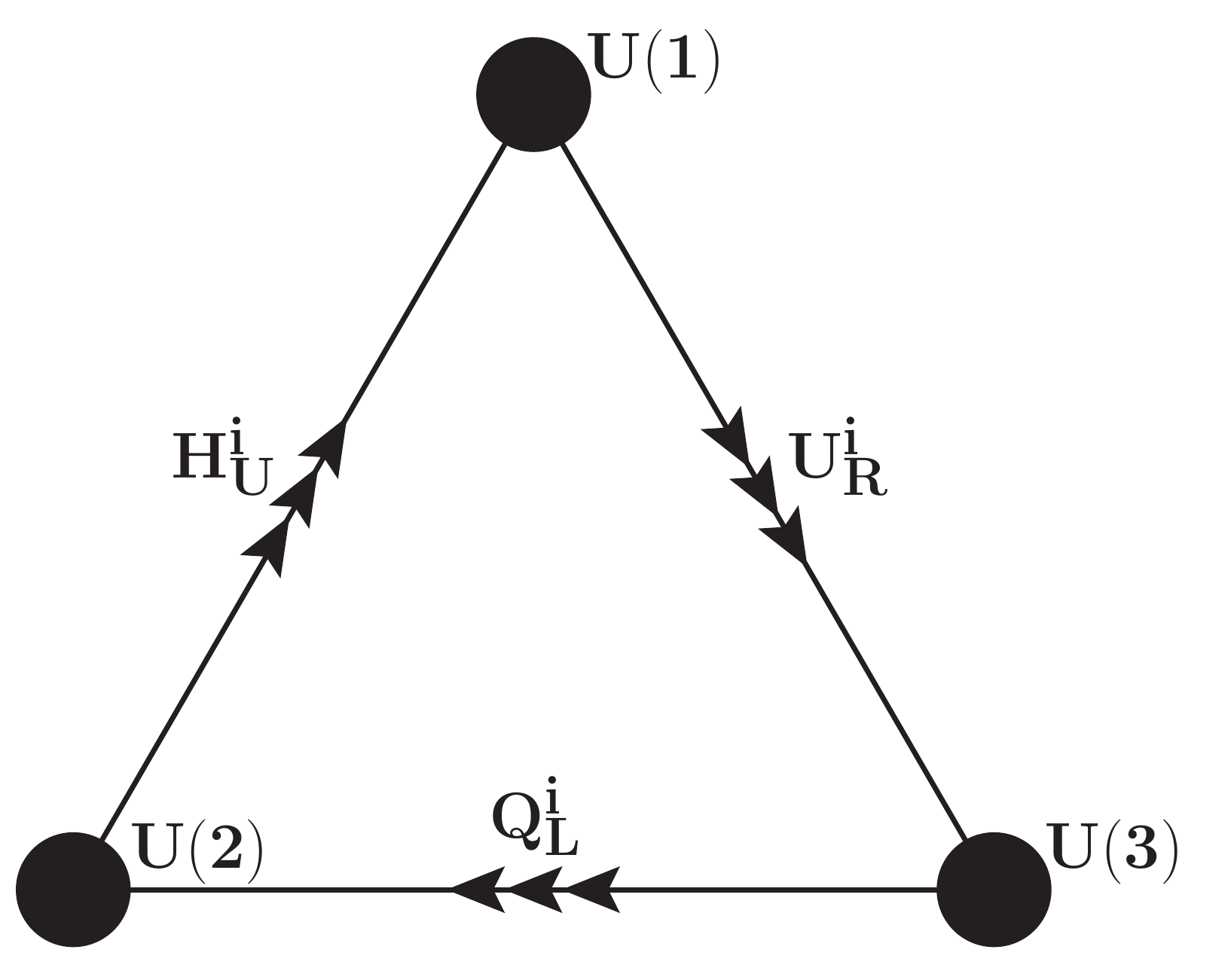}
\caption{Quiver diagram for D3-branes probing a $\czt$ singularity.}
\label{quiver}
\end{center}
\end{figure}

    The massless spectrum of D3-branes probing a $\czt$ singularity  is summarised in  Figure \ref{quiver} by a quiver diagram; 
gauge groups are indicated by nodes and bifundamental chiral multiplets  are indicated by arrows between the
nodes. Each chiral multiplet has a threefold degeneracy. This provides the generation
index for the models in \cite{bottomup, realistic, toric}. As shown in the figure, the chiral multiplets are identified with
the left-handed quarks $(Q_L^i)$, the right-handed up quarks $(U_{R}^{i})$ and the up Higgs $(H_{u}^{i})$.
They interact via a Yukawa interaction 
\be
  \epsilon_{ijk}H^{i}_{u} Q^{j}_{L} U^{k}_{R}
\ee
invariant under the isometry group $SU(3) \times U(1)$ with $H^{i}_{U}$, $ Q^{i}_{L}$ and $U^{i}_{R}$
transforming in the fundamental.

In order to obtain the form of the leading operators which break the symmetry in the Yukawa interactions, recall that the product of three fundamentals  $(H^{i}_{U},  Q^{i}_{L}, U^{i}_{R})$
decomposes into irreducibles as 
\be
\label{smdecom}
     (1,0) \times (1,0) \times (1,0) \to (0,0) \op (1,1) \op (1,1) \op (3,0)
\ee
  In section \ref{secpiso}  we found that the metric fluctuations with the smallest scaling dimension  which break the $SU(3)$ symmetry (flavons vevs) transform as $(1,1)$ and $(2,2)$. The operators which break the symmetry in the open string sector are gauge singlets that can be formed from the product  of the flavon vevs  with  the irreducible representations that appear in the decomposition
(\ref{smdecom}). Such singlets arise from the  product  \footnote{The are no singlets in the 
tensor products $(1,1) \times (3,0)$, $(2,2) \times (1,1)$ and $(2,2) \times (3,0)$.}
\be
   (1,1) \times (1,1) \to (0,0)  \op (1,1) \op(1,1) \op (2,2) \op (3,0)   
\ee
Thus once the bulk modes are integrated out the leading   symmetry breaking operators 
that can arise in the open string sector are given by the (1,1)s in \pref{smdecom}. These are
 \bea 
      {\lambda}^{\phantom{j} i}_{j} &=& \epsilon_{jmn} H^{(i}_{U} Q^{m)}_{L} U^{n}_{R},  \cr
      {\beta}^{\phantom{j} i}_{j} &=& \epsilon_{jmn} H_{U}^{m} Q_{L}^{n} U_{R}^{i}
       - {1 \over 3} \delta^{\phantom{j} i}_{j} \epsilon_{mnp} H_{U}^{m} Q_{L}^{n} U_{R}^{p}.
 \eea
The volume dependence of the strength of these operators is given by that of the flavon vev \pref{flv}, i.e.
${\cal{V}}^{-1/3}$.  It will be interesting to perform the complimentary analysis  of  a CFT computation involving three open string states
and a bulk KK mode in the $T^{6}/ \mathbb{Z}_{3}$ toriodal orbifold and infer the pattern of symmetry breaking in the open string sector from the amplitude.

\comments{
   We have used gauge invariance to enumerate the possible form of  operators that break the 
$SU(3)$ invariance of the Yukawa interactions; their presence can be checked by performing 
a CFT computation in the non-compact $\czt$ orbifold involving three open string modes and a graviton
\footnote{ While this computation will give the tensor structure of the interaction between the bulk modes and the open string states,
it will not reveal the volume dependence of the interaction and the fact that the mode in the $SU(3)$ representation (1,1) dominates; since these features are intrinsic to the compactified geometry.}.
We hope to perform this computation in the future.}

\section{Discussion}

   Central to our analysis has been  the assumption that at the asymptotic end of the cone all terms in the expansion \pref{expan} are
of equal magnitude. This assumption can be checked by exploiting the progress  in computing numerical metrics on Calabi-Yau manifolds \cite{Headrick, mrd}.
A  one parameter family of Ricci flat metrics on $K3$ was obtained in \cite{Headrick} by starting from the orbifold $\text{T}^{4}/ \mathbb{Z}_{2}$, and resolving the singular points to a finite size two sphere. In the large volume limit the local geometry near the resolutions is given by the Eguchi-Hanson geometry with corrections associated with finite volume effects. The numerical metric provides  a setting in which
one has  explicit knowledge of  the Fourier
coefficients $c_{I}$,  can be used to check the reliability of the assumptions on the size
of the Fourier coefficients.

    We have confined our attention to breaking of isometries as result of metric perturbations. A  compactification  with moduli stabilisation is going to have other fields such as fluxes and a non-trivial warp factor which can also lead to symmetry breaking. In the analysis of realistic models one  has to include these effects. Each mode would have an associated SU(3) representation. For instance the warp factor satisfies the scalar Laplacian equation in the extra dimensions and transforms with SO(6) Dynkin label ($l$,0,0) . Furthermore, as was emphasised in the context of brane inflation in \cite{holo2}, the leading physical effect can arise in second-order perturbation theory. Reference \cite{Heiden}
provides the back-reacted geometry of $\czt$ with the blow-up mode  stabilised by gaugino condensation on D7-branes. The geometry has a $SU(3)$ isometry.

\section{Conclusions}

  In this paper we have studied the symmetry breaking effects  that arise as a result 
  of compactification in local models on $\czt$. The ``power-law filtering''
  of modes    in cone-like geometries  combined with the requirement of
gauge invariance under the flavon group  allowed us to obtain the strength
and form of the leading operators responsible for the symmetry breaking. At large
volume we found a hierarchical separation in the breaking parameters for the 
$SU(3)$ and $U(1)$ symmetries.
  
   Although our focus has been on models on $\mathbb{C}^{3}/ \mathbb{Z}_{3}$, the approach is quite general and should be useful in understanding how symmetry breaking effects in  the bulk are communicated to the open string sector in local models. 
It will be interesting to carry out a similar analysis for models 
on complex cones over del Pezzo surfaces \footnote{These geometries have a $E_{n}$ global symmetry \cite{hidden}.} \cite{Verlinde,realistic,toric,dpt} and F-theory models \cite{fone, FthSym} and study
implications for  flavour structure.

\section*{Acknowledgments}

 I would like to thank Maciej Dunajski, Thomas Grimm, Liam McAllister, Fernando Quevedo and Ashoke Sen for very helpful communication and discussion.  AM is funded by the European Union under the Seventh Framework Programme (FP7) and the University of Cambridge. 

\appendix

\section*{Appendix}

\subsection*{Decomposition of tensor harmonics on ${\bf{S^{5}}}$ under ${\bf{SU(3)}} {\bf{\times}} {\bf{U(1)}}$ }
\label{appdecom}

 As discussed in section \ref{secpiso} tensor harmonics on the the five sphere are labelled by a single quantum
number  $l (\geq 2) $, with the highest weight state having $SO(6)$ Dynkin label $(l-2,2,2)$. In this appendix we
decompose the   $l=2,3$ and $4$ representations under $SU(3) \times U(1)$.
\begin{itemize}
\item $l=2$
\bea
\nonumber
   (0,2,2) \to &&(2,0)^{-4} \oplus (1,0)^{-2} \oplus (2,1)^{-2} \oplus \cr
    &&(0,0)^{0} \oplus  (1,1)^{0} \oplus (2,2)^{0} \op \cr 
     &&(1,2)^{+2} \oplus (0,1)^{+2} \oplus (0,2)^{+4}
\eea
\item $l=3$
\bea
\nonumber
   (1,2,2) \to &&(2,1)^{-5} \oplus (1,1)^{-3} \oplus (3,0)^{-3} \oplus (2,2)^{-3} \oplus  \cr && (0,1)^{-1} \op (2,0)^{-1} \op
   (1,2)^{-1} \op (3,1)^{-1} \oplus (2,3)^{-1} \op \cr
    &&(3,2)^{+1} \op (1,3)^{+1} \op (2,1)^{+1} \op (0,2)^{+1} \op (1,0)^{+1} \op \cr
    && (2,2)^{+3} \op (0,3)^{+3} \op (1,1)^{+3} \op (1,2)^{+5}
    \eea
\item $l$ = 4
\bea
\nonumber
(2,2,2) \to &&  (2,2)^{-6} \op (1,2)^{-4} \op  (3,1)^{-4} \op (2,3)^{-4} \op \cr 
&& (0,2)^{-2} \op (2,1)^{-2} \op (1,3)^{-2} \op (4,0)^{-2} \op (3,2)^{-2} \op (2,4)^{-2} \op \cr 
&&  (0,3)^0  \op (4,1)^0 \op  (1,1)^0 \op (2,2)^0 \op (3,3)^0 \op (1,4)^0 \op (3,0)^0 \op  \cr
&& (4,2)^{+2} \op  (2,3)^{+2} \op (0,4)^{+2} \op (3,1)^{+2} \op (1,2)^{+2} \op (2,0)^{+2} \op \cr
&& (3,2)^{4} \op (1,3)^{4} \op (2,1)^{4} \op (2,2)^6
\eea  
\end{itemize}
\comments{
\subsection*{group theory with dimensions}
\be
   (0,2,2)_{84} \to (2,0)_{6} \oplus (1,0)_{3} \oplus (2,1)_{15} \oplus (0,0)_{1} \oplus  (1,1)_{8} \oplus (2,2)_{27} \oplus (1,2)_{15}  \oplus (0,1)_{3} \oplus (0,2)_{6} 
\ee
\bea
\nonumber
   (1,2,2)_{300} \to &&(2,1)_{15} \oplus (1,1)_{8} \oplus (3,0)_{10} \oplus (2,2)_{27} \oplus (0,1)_3 \op (2,0)_{6} \op
   (1,2)_{15} \op (3,1)_{24} \oplus (2,3)_{42} \op \cr
    &&(3,2)_{42} \op (1,3)_{24} \op (2,1)_{15} \op (0,2)_6 \op (1,0)_3  \op (2,2)_ {27} \op (0,3)_{10} \op (1,1)_8 \op (1,2)_{15} 
\eea
\bea
\nonumber
(2,2,2)_{729} \to && (2,2)_{27} \op (1,2)_{15} \op  (3,1)_{24} \op (2,3)_{42} \op (0,2)_6 \op (2,1)_{15} \op (1,3)_{24} \op (4,0)_{15} \op (3,2)_{42} \op (2,4)_{60} \op \cr 
&&  (0,3)_{10}  \op (4,1)_{35}\op  (1,1)_8 \op (2,2)_{27} \op (3,3)_{64} \op (1,4)_{35} \op (3,0)_{10} \op  \cr
&& (4,2)_{60} \op  (2,3)_{42} \op (0,4)_{15} \op (3,1)_{24} \op (1,2)_{15} \op (2,0)_{6} \op (3,2)_{42} \op (1,3)_{24} \op (2,1)_{15} \op (2,2)_{27}
\eea
}

\end{document}